\documentclass[aps,twocolumn,showpacs,email,amssymb,nobibnotes]{revtex4-1}

\usepackage{graphicx,subfig,bm}

\usepackage{amsmath}
\usepackage{amsthm}
\usepackage{eucal}
\usepackage{amssymb}
\usepackage{mathrsfs}
\usepackage{float}

\begin{document}

\title{Modes of competition and the fitness of evolved populations}

\author{Tim Rogers$^{1}$ and Alan J.~McKane$^{2}$}

\affiliation{$^{1}$Centre for Networks and Collective Behaviour, Department of 
Mathematical Sciences, \\ 
University of Bath, Claverton Down, Bath, BA2 7AY, UK \\
$^{2}$Theoretical Physics Division, School of Physics and Astronomy, 
The University of Manchester, Manchester M13 9PL, UK}

\begin{abstract}
Competition between individuals drives the evolution of whole species. Although the fittest individuals survive the longest and produce the most offspring, in some circumstances the resulting species may not be optimally fit. Here, using theoretical analysis and stochastic simulations of a simple model ecology, we show how the mode of competition can profoundly affect the fitness of evolved species. When individuals compete directly with one another, the adaptive dynamics framework provides accurate predictions for the number and distribution of species, which occupy positions of maximal fitness. By contrast, if competition is mediated by the consumption of a common resource then demographic noise leads to the stabilization of species with near minimal fitness. 
\end{abstract}
\maketitle

\section{Introduction and model specification}
\label{sec:intro}

Evolution is often thought of as a process of optimization: beneficial mutations are accrued over generations as individuals adapted to best exploit their environment are selected for. Here, the meaning of `beneficial' is restricted to the sense of conferring a competitive advantage, allowing the mutant individual to produce more and more successful offspring than members of the general population. So it is that the basic mechanisms of mutation and competition together shape the characteristics of individuals and species. 

There have been many attempts to formalize these concepts in a mathematical setting, including dynamical models~\cite{Nowak2006} and comparisons to optimization theory~\cite{Grafen2002,Gyllenberg2011}. The theory of \textit{adaptive dynamics} is a popular framework for describing the evolution of phenotypes in a simplified `trait space'~\cite{Metz1996,Eshel1997,Geritz1997,Geritz1998,Diekmann2004,Geritz2004,Meszena2005,Waxman2005,Kisdi2010}. Adaptive dynamics is intended to describe the long-term behaviour of species in the limit of large population sizes, under the assumption that the occurrence of new mutations is vanishingly rare. Under these conditions, evolution follows a gradual course of speciation and optimization, until (typically) a configuration of species is found in which each occupies a position of maximal fitness. Situations exist, however, in which evolution appears to deterministically drive populations towards sub-optimal states or even to the point of extinction \cite{Gyllenberg2000,Gyllenberg2001}. Here we will show how stochastic effects arising from demographic noise can also drive the formation of sub-optimal species. 

Historically, the link between the interaction of individual organisms and the collective change of a species as a whole has been somewhat taken for granted, with demographic noise thought to play only a secondary role in the evolution of species. However, we have recently shown that even in very large populations stochasticity can have a powerful effect on the outcome of competitive processes~\cite{Rogers2012,Rogers2012a}, and it may be argued that these considerations can be important for large classes of organisms~\cite{Rossberg2013}. This work demonstrates the need to examine in detail the relationship between mutation rate, population size and competition between individual organisms. 

In this article, we explore the role played by the type of competition on the formation and fitness of species in a stochastic model ecology. The number and variety of models exploring the phenomenon of speciation is very large (for reviews see, for instance~\cite{Kirkpatrick2002,Gavrilets2004,Bolnick2007}), but here we are concerning specifically with those formulated within the context of adaptive dynamics, and more particularly models which are inherently stochastic. Even then, there have been many studies of stochasticity and instabilities in adaptive dynamics~\cite{Champagnat2001,Champagnat2006,Champagnat2007,Perthame2007,Claessen2007,Claessen2008,Meleard2009,Perthame2010,Lorz2011,Klebaner2011,Sagitov2013}, but here we will be especially interested in the differences that arise between direct predation, and indirect competition for a common resource. Numerical simulations of these processes suggest that in the first case, the population organizes into species of optimal fitness, and we are able to employ adaptive dynamics to make successful predictions about this process. With indirect competition, however, species form which are typically of near minimal fitness. This behaviour does not agree with the predictions of adaptive dynamics, and a full stochastic analysis is required. 

The paper is organised as follows. In the remainder of this section we give details of the model specification. In Section \ref{secII} the adaptive dynamics analysis of the model in the case of direct competition is described, while Section \ref{secIII} explores the role of demographic noise when competition is indirect. In the final section we discuss the implications of our work for evolutionary theory. 

The model specification is the same as that studied in~\cite{Rogers2012,Rossberg2013}, which we now recap. We note that our model is of a type which is expected to exactly conform to the predictions of adaptive dynamics, as established in \cite{Champagnat2006}. At time $t$ the state of the system consists of $N=N(t)$ organisms, with phenotypes having values $x_1\,,\,\ldots\,,\,x_N$ in a trait space. We choose this trait space to be the one-dimensional line segment $[-\pi,\pi)$ with periodic boundary conditions; this choice is made for mathematical convenience and to eliminate boundary effects. The dynamics of the model are as follows. Each organism reproduces asexually with rate one, with the phenotype of the offspring being that of the parent, plus a random mutation chosen from the normal distribution with mean zero and variance $\mu$. Organisms die as a result of competition; the death rate of an organism with phenotype $x$ is given by
\begin{equation}
d(x)=\frac{1}{K}\sum_{i=1}^{N} g(x-x_i)\,.
\end{equation}
Here the parameter $K$ controls the carrying capacity of the system and the competition kernel $g$ describes the strength of competition as a function of the difference in phenotype. We take $g$ to be positive, symmetric, and monotonically decreasing away from zero. The biologically relevant parameter regime describes large populations with very weak mutation effects. Mathematically this corresponds to the limit $K\to\infty$ and $\mu\to0$, with the product $\mu K$ being small. 

The state of the system at a given moment in time, $t$, is specified by the population density 
\begin{equation}
\phi(x,t)=\frac{1}{K}\sum_{i=1}^N\delta(x-x_i)\,.
\label{def_phi}
\end{equation}
Here the time dependence derives from the varying trait values $x_i$ and the number of organisms, but we will not make this dependence explicit to avoid clutter. Either on phenomenological grounds, or following the calculations of~\cite{Rogers2012}, we are able to write down a macroscopic description of the dynamics of $\phi$ in the limit $K\to\infty$,
\begin{equation}
\frac{\partial \phi(x,t)}{\partial t} = s(x,t)\phi(x,t)+\mu\,\nabla^2\phi(x,t)\,.
\label{macro_phi}
\end{equation}
The function $s(x,t)$ used here is the \emph{invasion fitness}, defined by
\begin{equation}
s(x,t)=1-\int^{\pi}_{-\pi}\,\phi(y,t)g(x-y)\,dy\,.
\label{invas_fit}
\end{equation}
The invasion fitness is so called because $s(x,t)$ describes the per-capita growth rate which would be experienced by a new species inserted at position $x$ in trait space at time $t$. The adaptive dynamics formalism is based on an analysis of this function, which is assumed to dominate the behaviour of (\ref{macro_phi}) when $\mu$ is small. 

We now go on to discuss separately the cases of direct and indirect competition.

\section{Direct competition}
\label{secII}
We begin by analyzing the behaviour of the model in the case that two organisms compete directly if their phenotypes are within some measure of similarity, and not at all otherwise. The closest biological analogies in this situation are cannibalism, and symmetric intra-guild predation~\cite{Polis1989} whereby two individuals may attack each other if they share some of the same prey. 

We assume that there is a finite interaction range $[-w,w]$ outside of which the competition kernel is zero. To obtain sensible predictions using adaptive dynamics, it is useful to make two further assumptions: (i) negative curvature at the origin, $g''(0)<0$, and (ii) that $g'$ and $g''$ are zero at the edges $\pm w$. Although these conditions appear to rule out several simple choices, such as a top-hat or truncated Gaussian, there is still considerable freedom, and it is possible to choose $g$ arbitrarily close to these examples. 

\begin{figure}
\begin{center}
\includegraphics[width=0.48\textwidth, trim=10 2 30 0, clip=true]{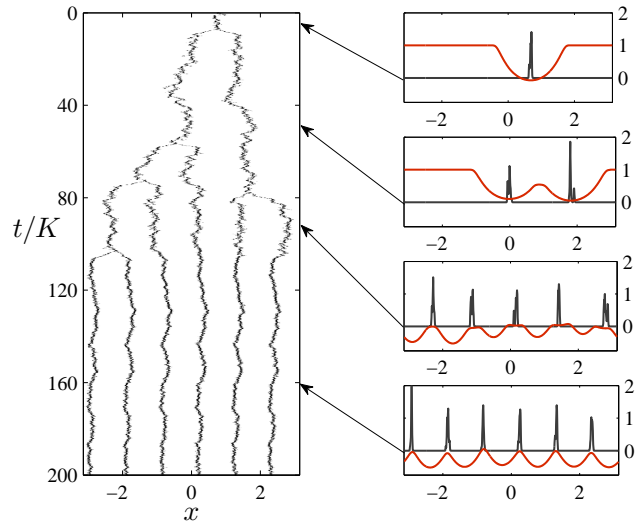}
\caption{Left: density map of a stochastic simulation of the direct competition model, with parameters $K=10^3$, $\mu=10^{-5}$, $w=1.2$ and kernel $g(x)\propto\exp\left\{((x/w)^2-1)^{-1}\right\}$. Right: snapshots of the population density $\phi$ (black) and invasion fitness $s$ (red) at various time points.}
\label{direct_fig}
\end{center}
\end{figure}

Stochastic simulations of the model with this choice of competition kernel are possible using the Gillespie algorithm~\cite{Gillespie1976,Gillespie1977}. As shown in Fig.~\ref{direct_fig}, the simulated population develops a number of distinct clusters of organisms, which may be interpreted as species. We will analyze the behaviour of these species using adaptive dynamics. It should be stressed that the following analysis holds for any $g$ with the above listed properties. 


The adaptive dynamics framework is intended to apply in the limit of almost faithful reproduction (that is, the diffusion term in Eq.~(\ref{macro_phi}) is ignored), where it is claimed that the basic properties of a model can be determined from an analysis of the invasion fitness \cite{Metz1996,Eshel1997,Geritz1997,Geritz1998,Diekmann2004,Geritz2004,Meszena2005}. In this regime, we may assume that the population is composed of $M$ distinct species, with all members of a species having the same phenotype. The population density $\phi$ may thus be decomposed as
\begin{equation}
\phi(x,t)=\sum_{\alpha=1}^M\delta(x-x_\alpha)\psi_\alpha(t)\,,
\label{delta_phi}
\end{equation}
where $x_\alpha$ and $\psi_\alpha$ respectively denote the phenotype and population size of species $\alpha$ (here and hereafter we use Greek indices for species, and Roman indices for individuals). The delta-function in Eq.~(\ref{delta_phi}) may be replaced by a distribution with some spread in trait space~\cite{Sasaki2011}, but we will not study this model variant here. 

To extract the value $\psi_\alpha$ from the population density $\phi$, we define a small interval, $I_\alpha$, which contains all individuals of phenotype $x_\alpha$, and no individuals of other phenotypes. Then using Eq.~(\ref{delta_phi}),
\begin{equation}
\int_{I_\alpha}\phi(x,t)\,dx = \sum_{\beta=1}^M\,\psi_\beta(t)\,\int_{I_\alpha}\delta(x-x_\beta)\,dx = \psi_\alpha(t)\,.
\end{equation}
Putting $\mu=0$ in Eq.~(\ref{macro_phi}) and integrating over $I_\alpha$ we obtain
\begin{equation}
\frac{d\psi_\alpha(t)}{dt} = s(x_\alpha,t)\,\psi_\alpha(t)\,,
\label{psi_eq}
\end{equation}
where the invasion fitness now takes the form
\begin{equation}
s(x,t)=1-\sum_{\beta=1}^M g(x-x_\beta)\,\psi_{\beta}(t)\,.
\label{s_psi}
\end{equation}
According to adaptive dynamics, the propensity for species $\alpha$ to adapt is captured by the \emph{local fitness gradient} at $x_\alpha$, 
\begin{equation}
s'(x_\alpha,t)=-\sum_{\beta=1}^M g'(x_\alpha-x_\beta)\,\psi_{\beta}(t)\,,
\label{lfg}
\end{equation}
where the prime denotes differentiation with respect to the first argument, $x$. In what follows we will drop the explicit $t$-dependence of $s$ to avoid clutter.

For as long as $s'(x_\alpha)\neq0$, species $\alpha$ is expected move slowly uphill in $s$, as a result of many small mutation events. If it has reached a state where $s'(x_\alpha)=0$ then there is no preferred direction of movement. There are now two possibilities. If $s''(x_\alpha)<0$ then the species resides at a local maximum of the fitness landscape, and is considered to be stable. Alternatively, if $s''(x_\alpha)>0$, then the species is located at a fitness minimum and would be expected to speciate, that is, the population will bifurcate into two sub-populations with phenotypes just either side of that of the ancestor species. These daughter species are then likely to experience a non-zero fitness gradient and will begin to move.

The long-term prediction of adaptive dynamics is that all species will undergo a process of continuing evolution and bifurcation, until every species resides at a local fitness maximum. The final configuration is referred to as an evolutionarily stable strategy. In the remainder of this section, we will follow the adaptive dynamics calculation in detail to derive a prediction for the final state of the model. As we will explain, we expect to find species evenly spread around trait space, with their number given by a certain stability condition.  

We begin with the case where there is only one initial species ($M=1$). From Eq.~(\ref{s_psi}), the invasion fitness is simply given by $s(x)=1-g(x-x_1)\psi_1(t)$. At large times, we would expect $\psi_1(t)$ to approach a constant value, which from Eq.~(\ref{psi_eq}) implies that $s(x_1)=0$. From this we can predict the quasi-stable population size: $\psi_1=1/g(0)$. In the absence of other species there is no advantage to be gained by a change in phenotype; the local fitness gradient at $x_1$ is from Eq.~(\ref{lfg}) $s'(x_1)=-g'(0)/g(0)=0$, since $g$ is even. To find the stability of the species, we differentiate Eq.~(\ref{s_psi}) twice with respect to $x$ and set $x=x_1$ to obtain $s''(x_1)=-g''(0)/g(0)$. Since, by assumption, $g''(0)<0$, $s''(x_1)>0$, and
the single initial species is always unstable and prone to bifurcation. 

Following the bifurcation of the initial species, we now have two species of approximately equal sizes and nearby phenotypes. For $M=2$ species, the fitness gradient at $x_1$ is given by
\begin{equation}
s'(x_1)=-g'(x_1-x_2)\,\psi_2(t)\,.
\end{equation}
Because $g$ is symmetric and decreasing with $|x|$, we know that its derivative changes from positive to negative at the origin. Thus $s'(x_1)\geq0$ if $x_1\geq x_2$ and $s'(x_1)\leq0$ if $x_1\leq x_2$. The results for $s'(x_2)$ can be found by interchange of the indices $1$ and $2$. Therefore, if the species move uphill in fitness --- as we would expect --- they will be moving apart in phenotypic space. 

Once the species have moved a distance $w$ apart, they will no longer compete and the phenotypes will cease to evolve. At this stage, we see from Eq.~(\ref{psi_eq}), that $s(x_1)=0$ and $s(x_2)=0$, which using Eq.~(\ref{s_psi}) gives
\begin{eqnarray}
g(0)\psi_1 + g(w)\psi_2 = 1, \ \ \  g(w)\psi_1 + g(0)\psi_2 = 1,
\end{eqnarray}
using $g(x_1 - x_2) = g(x_2 - x_1) = g(w)$. However, $g(x)$ (and all its derivatives) are zero for $|x| \geq w$, and therefore we have once again that $\psi_{\alpha}= 1/g(0)$, $\alpha=1,2$. Proceeding as in the $M=1$ case we can again show that $s'(x_{\alpha})=0$ and $s''(x_{\alpha}) = - g''(0)/g(0) > 0$, for $\alpha=1,2$, and so the $M=2$ configuration is also unstable.

Although both species are now prone to bifurcation, it is extremely unlikely that they will do so at the same time. Instead, we expect that one will speciate before the other, leading to a situation with three species; two close together and one at distance just less than $w$. Repeating the arguments for the two species situation, we expect that the three species will mutually move apart. 

For an arbitrary number of species $M\geq3$, it is necessary to start to consider the possibility that not all species will be able to get to a distance more than $w$ apart, since they may not be able to fit into the interval $[-\pi,\pi)$. More specifically, if $Mw < 2\pi$, then it is possible for the species to achieve a separation of at least $w$ between any pair. In this case, by an analogous set of arguments to those used above, all the species are unstable, and so one will bifurcate, leading to an increase in the number of species to $M+1$.

We are therefore led to consider the case where $Mw \geq 2\pi$. The best that can be achieved now is for the species to move as far away from others as possible, leading to the species being evenly spaced around phenotypic space. In this configuration the species may be stable, or they may not; further analysis is required. 

We begin by finding the long time properties of the configuration consisting of $M$ equally spaced species. As before, from Eq.~(\ref{psi_eq}) we have $s(x_\alpha)=0$ and from Eq.~(\ref{s_psi})
\begin{equation}
\sum^{M}_{\beta = 1} g(x_\alpha - x_\beta)\,\psi_\beta = 1, \ \ \alpha=1,\ldots,M.
\label{M_spaced}
\end{equation}
We can solve for $\psi_\alpha$ by regarding $g$ as a matrix. One finds that $\psi_\alpha$ has the same value for all $\alpha$, and so all species have the same population size. By translational invariance we can assume that one species is located at the origin, which leads to
\begin{equation}
\psi_\alpha=\psi=\frac{1}{\sum_{\beta=1}^{M} g(x_\beta - x_\alpha)} =
\frac{1}{\sum_{\beta=1}^M g(x_\beta)},
\label{equal_psi}
\end{equation}
which clearly satisfies Eq.~(\ref{M_spaced}). In addition, from Eq.~(\ref{lfg})
\begin{equation}
s'(x_\alpha)=-\psi\sum_{\beta=1}^M g'(x_\beta)\,.
\label{s_prime_M}
\end{equation}

To evaluate the sum in Eq.~(\ref{s_prime_M}) we look at the cases $M$ odd and $M$ even separately. First suppose that $M$ is odd. One species is at the origin, but $g'(0)=0$, and so it does not contribute to the sum. The other $M-1$ species  will occur in pairs at $x <0$ and $x>0$, equidistant from the origin. Since $g'(x) = - g'(x)$ these will cancel in pairs. Therefore the sum is zero. If $M$ is even, the same argument goes through, except there will be an additional species at $-\pi$, which will also not contribute since $g'(-\pi)=0$. So the sum is again zero. We conclude that the configuration with $M$ equally spaced species with equal populations is a fixed point of the adaptive dynamics.

It only remains to determine the stability of this configuration. To do this, we need to calculate the sign of
\begin{equation}
s''(x_\alpha)=-\psi\sum_{\beta=1}^M g''(x_\beta)\,.
\label{stability_M}
\end{equation}
It will be useful later to introduce an equivalent measure, the population average of the fitness curvature
\begin{equation}
Q=\frac{1}{N}\sum_{i=1}^N s''(x_i)\,.
\label{def_Q}
\end{equation}
Note that the sum here is over individuals, not species. Since from Eq.~(\ref{stability_M}) the second derivative of $s$ is independent of species, in the present case we simply have $Q=s''(x_\alpha)$. If $Q<0$ the $M$-species configuration is stable and if $Q>0$ it is not. 

The adaptive dynamics story for this model then is as follows. From an initial monomorphic population, we expect species to bifurcate and move apart (either to distance $w$ or as much as possible) until finding a configuration of $M$ stable species for which $Q<0$. Simulation of the stochastic version of the model for large $K$ generally shows good agreement with these predictions, as shown in Fig.~\ref{direct_fig}.

\section{Indirect competition}
\label{secIII}
We move on now to consider the situation that organisms do not predate on each other directly, but rather they consume a common resource. This is the more common mode of competition in real ecosystems. We take the function $g$ from the previous section to now denote the range and amount of resources an organism can exploit, so that the relative competition two organisms experience is related to the overlap of their respective resource-use curves. In Appendix~\ref{apph} we show that the resulting competition kernel is given by
\begin{equation}
h(x)=\frac{1}{2\pi}\int^{\pi}_{-\pi} g(x-y)g(y)\,dy\,.
\label{def_h}
\end{equation}
Note that the spatial extent of the kernel $h$ is twice that of $g$, although this does not in fact have a big effect on the dynamics. More importantly, the \textit{shape} of the kernel has changed: $h$ is necessarily smoother and shallower than $g$. 

The adaptive dynamics calculation for the model with this competition kernel proceeds along the same lines as in the previous section, with $g$ replaced everywhere by $h$. In fact, the same conclusion holds: species will continue to move apart and to speciate for as long as $Q$ is positive. What is different, however, is that for this choice of competition kernel, $Q$ is \emph{always} positive. To see this, we pass to Fourier space, writing $\phi_k$, $g_k$ and $h_k$ for the Fourier modes of $\phi(x)$, $g(x)$ and $h(x)$, respectively. These are defined as follows:
\begin{equation}
\phi(x,t) = \frac{1}{2\pi}\,\sum_{k}\,\phi_{k}(t) e^{ikx},
\label{FT_defn}
\end{equation} 
where $k=0, \pm 1, \pm 2,\ldots$ and 
\begin{equation}
\phi_k(t) = \int^{\pi}_{-\pi} \phi(x,t) e^{-ikx}\,dx\,.
\label{FT_inverse}
\end{equation} 

To obtain an expression for $Q$ which we can analyze, we begin by differentiating Eq.~(\ref{invas_fit}) twice with respect to $x$ to find
\[
s''(x)= -\int^{\pi}_{-\pi}\,\phi(y,t)h''(x-y)\,dy\,,
\]
where we are now using $h$ as the competition kernel rather than $g$. Multiplying by $\phi(x,t)$ and integrating over $x$ gives
\begin{eqnarray}
&-& \int^{\pi}_{-\pi} dx\,\int^{\pi}_{-\pi} dy\,\phi(x,t) h''(x-y) \phi(y,t)
\nonumber \\
&=& \int^{\pi}_{-\pi} s''(x)\,\phi(x,t)\,dx = \frac{1}{K}\,\sum^{N}_{i=1}\,
s''(x_i),
\label{intermediate}
\end{eqnarray}
where we have used the definition (\ref{def_phi}) in the last line. Therefore from the definition of $Q$, and suppressing the time dependence,
\begin{equation}
\begin{split}
&Q=\frac{1}{N}\sum_n s''(x_n)\\
&=-\frac{K}{N}\,\int^{\pi}_{-\pi} dx\,\int^{\pi}_{-\pi} dy\,\phi(x)h''(x-y)\phi(y)\\
&=\frac{K}{N}\left(\frac{1}{2\pi}\right)^{\hspace{-1pt}3} \hspace{-3pt}\sum_{n,m,k} \phi_n\phi_m k^2h_k \int\hspace{-5pt}\int e^{i(n+k)x+i(m-k)y}\,dx\,dy\\
&=\frac{K}{N}\sum_k\,\frac{k^2 h_k}{2\pi} \phi_k\phi_{-k}\,.
\end{split}
\label{Qfour}
\end{equation}
Since $\phi(x)$ is a real-valued function, we know that $\phi_k\phi_{-k}=|\phi_k|^2$. In addition, from Eq.~(\ref{def_h}) we find the relationship $h_k=g_k^2/2\pi$, and from Eq.~(\ref{phi_zero}) in Appendix~\ref{appS} that $\phi_0 = N/K$. Therefore
\begin{equation}
Q = \frac{1}{\phi_0}\,\sum_k \left(\frac{k g_k}{2\pi}\right)^{\hspace{-1pt}2}
\left| \phi_k \right|^2\,,
\label{Qone}
\end{equation}
and thus $Q$ is expressed as a sum of positive terms. 

So for a competition kernel of the form (\ref{def_h}), adaptive dynamics predicts a never-ending process of spreading and subdivision of species, with the eventual outcome being a uniformly populated trait space. Indeed, in the case of a uniform population density $\phi(x)\equiv 1/2\pi$ we have from Eq.~(\ref{FT_inverse}), $\phi_k=\delta_{k,0}$ and from Eq.~(\ref{Qone}) we see that the fitness curvature achieves its theoretical minimum value of $Q=0$. 


\begin{figure}
\begin{center}
\includegraphics[width=0.48\textwidth, trim=14 12 30 0, clip=true]{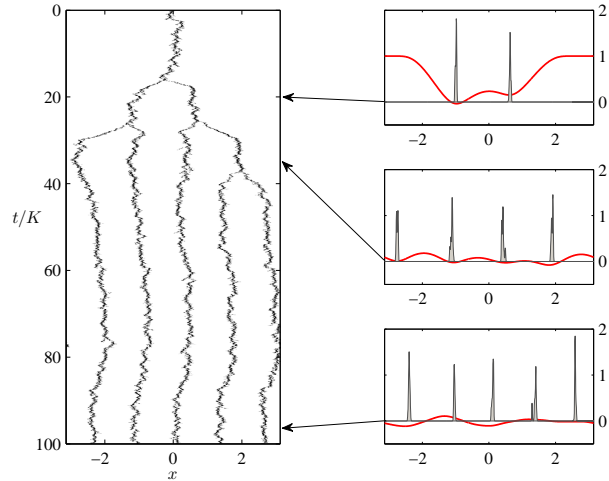}
\caption{Left: density map of a stochastic simulation of the indirect competition model, with parameters $K=10^3$, $\mu=10^{-5}$, $w=1$, and the underlying kernel $g$ is the same as in Figure 1. Right: snapshots of the population density $\phi$ (black) and invasion fitness $s$ (red) at various time points. Note that the five apparently stable species do not occupy positions of maximal fitness; in fact at least two are near fitness minima.}
\label{indirect_fig}
\end{center}
\end{figure}

\begin{figure}
\begin{center}
\includegraphics[width=0.45\textwidth, trim=0 0 0 0, clip=true]{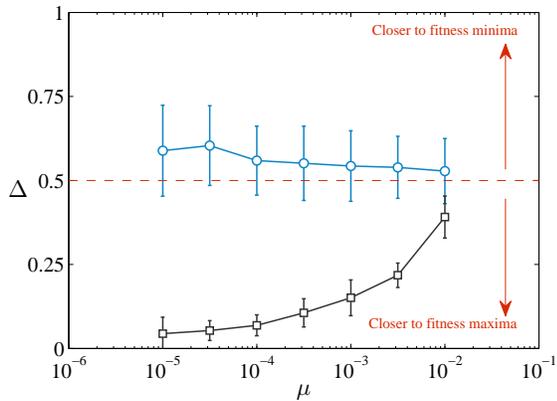}
\caption{Relative distance to optimality $\Delta$ (see text for explanation) as a function of mutation rate $\mu$. Average over 100 samples with error-bars for standard deviation) at time $t=10^6$ for populations developing under direct (black squares) and indirect (blue circles) competition. Parameters used were the same as figures \ref{direct_fig} and \ref{indirect_fig}. The red dashed line marks the halfway point; values above this line imply species that are closer to fitness minima than maxima.}
\label{fitness_comparison}
\end{center}
\end{figure}

In stochastic simulations, however, the system does not achieve a uniform distribution. Instead, we observe the creation of a finite number of well-separated species of approximately equal populations: see Fig.~\ref{indirect_fig} for an example. Surprisingly, these species tend to occupy areas of fitness space which are below average, often being located near fitness minima. A simple test of this observation is to measure the distance from the nearest local fitness maximum and minimum. For each organism $i$ we measure the distance $\Delta_i^+$ to the nearest local fitness maximum, and $\Delta_i^-$ to the nearest fitness minimum, then compute the relative distance from optimality $\Delta_i=\Delta_i^+/(\Delta_i^++\Delta_i^-)$. Averaging this over the population ($\Delta=\frac{1}{N}\sum_i\Delta_i$) gives an indication whether the species that have formed occupy favorable locations in the fitness landscape. In Fig.~\ref{fitness_comparison} we show how this quantity varies with $\mu$ for both direct and indirect competition. In the case of direct competition we see a straightforward relationship with $\Delta$ starting very small and increasing as $\mu$ is increased. The organisms are arranged in clusters centered at fitness maxima with width controlled by $\mu$; reducing mutation rate therefore enhances the overall fitness of the population by reducing the ``noise'' around the peak in the fitness landscape. For the case indirect competition there is no such relationship: for the whole range of $\mu$ tested we find species that are on average closer to fitness minima than maxima. 

As well as this simulation test we are able to develop theoretical predictions for another quantity that is proxy to fitness. We define the `invasibility' of a population by comparing the average fitness of extant organisms with the fitness of an invader introduced with a randomly chosen phenotype. Mathematically this is expressed as 
\begin{equation}
S=\frac{1}{2\pi}\int^{\pi}_{-\pi} s(x)\,dx-\frac{1}{N}\sum_{i=1}^Ns(x_i)\,.
\label{def_S}
\end{equation}
The first term here is the fitness of a randomly placed invader, the second is the average fitness of the existing population. If $S\leq0$ then an invading individual is typically no fitter than the members of the resident population; if $S>0$ then a random invader will be expected to out-compete the other organisms, meaning that the population is susceptible to invasions. In Appendix~\ref{appS} we show that 
\begin{equation}
S=\frac{1}{\phi_0}\sum_{k\neq0}\left(\frac{g_k}{2\pi}\right)^2|\phi_k|^2\geq 0\,,
\label{S_result}
\end{equation}
meaning that any non-uniform distribution will be susceptible to invasion. 

The species observed in Fig.~\ref{indirect_fig} are unfit: they are susceptible to invasion and they occupy locations which have positive fitness curvature. To understand the mechanism responsible for maintaining these unfit species, we must examine the model in greater detail, taking into account stochastic fluctuations in the population density. This requires working at large, but finite, carrying capacity.

Following ~\cite{Rogers2012}, it is possible to derive a mesoscopic description of the system in terms of a stochastic differential equation 
\begin{equation}
\frac{\partial \phi(x,t)}{\partial t} = s(x,t)\phi(x,t)+\mu\,\nabla^2\phi(x,t)+ 
\frac{\eta(x,t)}{\sqrt{K}}\,,
\label{SDE}
\end{equation}
where $\eta(x,t)$ is a Gaussian white noise with zero mean and a correlator which is given in Appendix~\ref{appFPE}. 

Our previous analysis suggested that the uniform population density should be an attractive fixed-point for the system, although simulations have shown otherwise. Taking the uniform population density as a starting point (that is, $\phi(x,0)=1/2\pi$), we will examine the growth of stochastic deviations away from this state by making the change of variables 
\begin{equation}
\zeta(x,t) = \phi(x,t) - \frac{1}{2\pi}.
\label{def_zeta}
\end{equation}
This analysis will apply to small times $t$ when $\zeta(x,t)$ is assumed to also be small, corresponding to the \textit{early-onset} regime describing the emergence of species clusters from a uniform density. Inserting (\ref{def_zeta}) into (\ref{SDE}) yields
\begin{eqnarray}
& & \frac{\partial \zeta(x,t)}{\partial t} = \mu\nabla^2\zeta(x,t)
- \frac{1}{2\pi}\,\int^{\pi}_{-\pi} h(x-y) \zeta(y,t)\,dy \nonumber \\ 
& & -\int^{\pi}_{-\pi} \zeta(x,t) h(x-y) \zeta(y,t)\,dy + 
\frac{\eta(x,t)}{\sqrt{K}}\,.
\label{SDE_zeta}
\end{eqnarray}

As before, the analysis is facilitated by moving to Fourier space. Defining $\zeta_{k}(t)$ and $\eta_{k}(t)$ to be the Fourier coefficients of $\zeta(x,t)$ and $\eta(x,t)$ respectively, analogously to Eq.~(\ref{FT_defn}), the Fourier transform of Eq.~(\ref{SDE_zeta}) is
\begin{equation}
\frac{d \zeta_k}{dt} = -\left(\mu k^2+\frac{h_k}{2\pi}\right)\zeta_k
- \frac{1}{2\pi}\sum_l h_l \zeta_l\zeta_{k-l} + \frac{\eta_k}{\sqrt{K}}\,.
\label{SDE_FT}
\end{equation}
From Eqs.~(\ref{Qone}) and (\ref{S_result}), we see that to determine the average of the stability measures, $Q$ and $S$, we need to know the average value of $|\phi_k|^2$. We therefore seek equations for the average values of $\zeta_k$ and $|\zeta_k|^2$. The first may simply be obtained by taking the average of Eq.~(\ref{SDE_FT}) remembering that the noise $\zeta_k(t)$ has zero mean:
\begin{equation}
\frac{d}{dt}\langle \zeta_k\rangle=-\left(\mu k^2+\frac{h_k}{2\pi}\right)\langle \zeta_k\rangle-\frac{1}{2\pi}\sum_l h_l\langle \zeta_l\zeta_{k-l}\rangle\,.
\label{m1}
\end{equation}
It is more straightforward to obtain an analogous expression for the time evolution of the product of $\zeta_{k}\zeta_{l}$ within the Fokker-Planck formalism. Using the results given in Appendix~\ref{appFPE} one can show for any pair of integers $k$ and $l$ that
\begin{equation}
\begin{split}
\frac{d}{dt}\langle \zeta_k\zeta_l\rangle=&-\left(\mu (k^2+l^2)+\frac{h_k+h_l}{2\pi}\right)\langle \zeta_k\zeta_l\rangle\\
&-\frac{1}{2\pi}\sum_m h_m\Big(\langle \zeta_m\zeta_{k-m}\zeta_l\rangle+\langle \zeta_m\zeta_{l-m}\zeta_k\rangle\Big)\\
&+\frac{1}{K\pi}+\frac{1}{K}\left(2-\mu (k+l)^2+\frac{h_{k+l}}{2\pi}\right)\langle \zeta_{k+l}\rangle\\
&+\frac{1}{2\pi K}\sum_m h_m\langle \zeta_m\zeta_{k+l-m}\rangle\,.
\end{split}
\label{m2}
\end{equation}

As is frequently the case, the hierarchy of moment equations is not closed: the
equation for the rate of change of $\langle \zeta_k \rangle$ depends on $\langle \zeta_k\zeta_{k-l}\rangle$, that for the rate of change of $\langle \zeta_k\zeta_l\rangle$ depends on $\langle \zeta_m\zeta_{l-m}\zeta_k\rangle$, and that for rate of change of $n^{\textrm{th}}$-order moments will depend on those of order $n+1$. To make progress we consider the early-onset behaviour as $\zeta(x,t)$ departs from zero. In this regime it is reasonable to assume that the third-order moments are sufficiently small that we may put $\langle \zeta_k\zeta_l\zeta_m\rangle\approx 0$. Returning to Eqs.~(\ref{m1}) and (\ref{m2}), we obtain the linear system for the infinite set of modes $|\zeta_k|^2$ and $\zeta_0$:
\begin{equation}
\frac{d}{dt}\langle \zeta_0\rangle=-\langle \zeta_0\rangle-\frac{1}{2\pi}\sum_k h_k\langle |\zeta_k|^2\rangle\,,
\label{zeta_0_av}
\end{equation}
\begin{eqnarray}
\frac{d}{dt}\langle |\zeta_k|^2\rangle&=&-\left(2\mu k^2+\frac{h_k}{\pi}\right)\langle |\zeta_k|^2\rangle \nonumber \\
&+&\frac{1}{K\pi}+\frac{3}{K}\langle \zeta_0\rangle+\frac{1}{2\pi K}\sum_l h_l\langle |\zeta_l|^2\rangle\,.
\label{zeta_sq_av}
\end{eqnarray}


\begin{figure}
\begin{center}
\includegraphics[width=0.48\textwidth, trim=18 0 25 0]{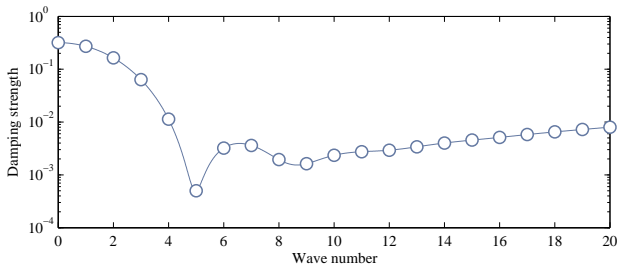} \\
\caption{Damping strength $2\mu k^2+h_k/\pi$ for different Fourier coefficients of the population density, with parameters the same as those of Fig.~\ref{indirect_fig}. }
\label{WD_fig}
\end{center}
\end{figure}


The first term on the right hand side of (\ref{zeta_sq_av}) describes a deterministic damping force, acting to reduce the amplitude of fluctuations. The remaining terms (of order $1/K$) capture the excitation due to demographic noise, which is independent of wave number. 

In terms of Fourier coefficients, the formation of $k$ species clusters in the population would be indicated by an excitation of wave number $k$. It is thus possible to predict the number of species that will form by computing the damping strength $2\mu k^2+h_k/\pi$ for different $k$. The wave number with the smallest damping will experience the largest noise-induced excitation. Fig.~\ref{WD_fig} shows the strength of damping for different values of $k$, with model parameters the same as those of Fig.~\ref{indirect_fig}. A clear minimum is achieved at $k=5$, correctly predicting the formation of five species. 


\begin{figure}
\begin{center}
\includegraphics[width=0.48\textwidth, trim=18 10 25 0]{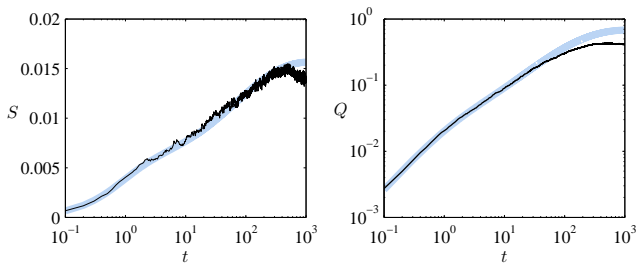} \\
\caption{The growth of invasibility ($S$, left) and mean fitness curvature ($Q$, right) in time, starting from an initially homogeneously distributed population, with parameters the same as those of Fig.~\ref{indirect_fig}. Thick lines represent the theoretical prediction (obtained by numerically integrating Eqs.~(\ref{zeta_0_av}) and (\ref{zeta_sq_av}) for the first 1000 wave modes), while thin lines show simulation results averaged over 200 samples.}
\label{Q_fig}
\end{center}
\end{figure}


Using Eqs.~(\ref{zeta_0_av}) and (\ref{zeta_sq_av}) we are also able to predict the evolution of the fitness measures $S$ and $Q$ for these noise-induced species. This is achieved by solving for the trajectories of $\langle|\zeta_k|^2\rangle$ and $\langle \zeta_0\rangle$, deducing the values $\langle |\phi_k|^2\rangle$, and inserting into Eqs.~(\ref{Qone}) and (\ref{S_result}). As the system is linear, an explicit expression for the solution could in principle be written down. In practice, however, it is more convenient to truncate after a suitably large $k_{\max}$ and numerically integrate the resulting finite system. In Fig.~\ref{Q_fig} we show a comparison between the simulated and theoretical values for $S$ and $Q$. In both cases the theory provides a successful description of the growth of these measures at early times, with deviations beginning to appear later. Since these measures are minimized by optimally fit species, the growth seen here indicates that the population as a whole is actually becoming less well suited to its environment. 

\section{Conclusions}

The analyses presented here have revealed the evolutionary consequences arising from different modes of competition. We considered a simple stochastic model of an evolving population under two different choices of competition kernel, one corresponding to \emph{direct} competition (e.g. intraguild predation, cannibalism), the other arising from \emph{indirect} competition induced by the need to consume a common resource. 

In the first case we found that direct competition leads eventually to the formation of a number of distinct species able to permanently coexist in an evolutionary stable situation. The number and distribution of species can be reliably predicted using the methods of adaptive dynamics. The evolutionary mechanisms at work in this regime are relatively straightforward: species migrate slowly `uphill' in the fitness landscape, with those located at local minima being prone to speciation and those at local maxima being stable. The final state of the population is one in which all species reside at local maxima of the fitness landscape; in this sense the process of evolution as reached an optimal configuration. 

The behaviour of the population evolving under indirect competition is more surprising. Snapshots of stochastic simulations of the system initially appear very similar to those in the case of direct competition, with the formation of several distinct species at approximately equidistant locations. Investigation of the fitness landscape, however, reveals that these species occupy locations of near minimal fitness. It appears that the usual adaptive dynamics argument employed in Section \ref{secII} --- that populations near a fitness minimum would be expected to speciate --- is no longer applicable. 

At a technical level, this phenomenon is just an example of the way that stochastic effects can modify deterministic dynamics, to give an effective dynamics which may differ considerably from the original form. For instance, new effective extrema may be formed through the effects of noise~\cite{Biancalani2014}, as well as the effective location of extrema being moved. In the present case, the effect is more closely related to the situation found in some kinds of stochastic pattern formation~\cite{Butler2009,McKane2013}, where the region of stability of patterns can be increased by the stochastic effects giving a new effective condition for stability which differs from that found through a purely deterministic analysis.

\begin{figure}
\begin{center}
\includegraphics[width=0.48\textwidth, trim=0 0 0 0]{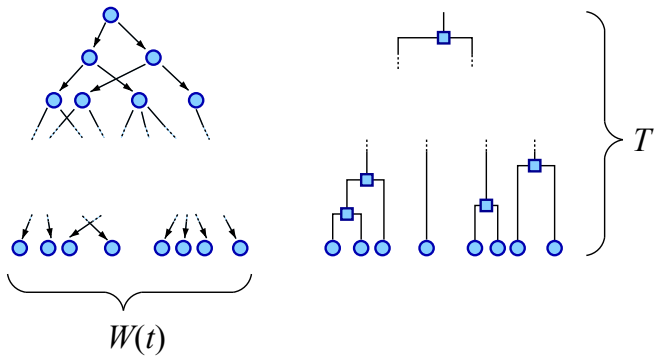} \\
\caption{Illustration of (a) the increasing diversity of descendants of a single initial organism, and (b) the typical time since a given population shared a single common ancestor.}
\label{explanation_fig}
\end{center}
\end{figure}

 In the present ecological context, a deeper intuition can be gained by considering the speed with which a species can explore phenotype space. As a result of many small mutations, the descendants of a single initial organism come to occupy a typical width $W(t)$ of trait space, which increases as a function of time, see Figure~\ref{explanation_fig}(a). Conversely, the common ancestors of a population can be traced back to form the phylogenetic tree; eventually leading to a single common ancestor at some time $T$ in the past, see Figure~\ref{explanation_fig}(b). A typical mature species will occupy an area of size around $W(T)$. If this value is too small in comparison to the width of the competition kernel, then species will be unable to bifurcate as they cannot explore a new areas of trait space fast enough. This phenomenon is responsible for the spontaneous speciation seen in \cite{Rogers2012,Rossberg2013}, and has recently been observed for very small population sizes in another model of adaptive dynamics \cite{Wakano2013}.


The work presented here can be extended and generalized in several different ways. There has been some progress made in the analysis of more complex models, showing that in some cases they may be shown to be equivalent to a generic adaptive dynamics model, but only in the case of deterministic dynamics~\cite{Doebeli2013}. It would be interesting to study if stochastic models also show generic features. We have also made some assumptions which warrant further investigation. For instance, we assumed periodic boundary conditions in the trait space, not of course because we expect the trait to be periodic, but to ensure analytic tractability. Previous authors have used this assumption, for instance Scheffer and van Nes~\cite{Scheffer2006}, who note that their result is robust against change of the niche axis from circular (``infinite periodic’'') to finite linear. A more realistic way of dealing with the boundaries might be to put a fitness penalty into the model to ensure low population densities in their vicinity. We have used this approach in this past in a related model~\cite{Rossberg2013}, as well as studying a model in which trait boundaries were entirely absent~\cite{Rogers2012a}. From our experience in these cases, we do not believe that the precise prescription that is used in dealing with the boundaries is important, but this needs to be studied further. More generally, we hope that the ideas we have presented here will encourage other researchers to probe the effects of demographic stochasticity on the theory of adaptive dynamics.

\section*{Acknowledgements}
We thank Axel Rossberg for valuable discussions. TR gratefully acknowledges the support of the Royal Society.

\begin{appendix}

\section{Derivation of $h$}
\label{apph}

In the main text we asserted that when organisms compete for a common resource, the effective interaction between individuals is described by the competition kernel
\begin{equation}
h(x)=\frac{1}{2\pi}\int^{\pi}_{-\pi} g(x-y)g(y)\,dy\,.
\label{def_h2}
\end{equation}
Here we motivate this definition. 

Suppose that the organisms depend on certain resources to live, and that position in trait space corresponds to resource type. We use the function $g(x-y)$ to now describe the efficiency with which an organism whose `favorite' resource is of type $x$ can consume resources of a different type $y$. The death rate for organisms of type $x$ can then be written as
\begin{equation}
d(x)=d_0-\gamma\int_{-\pi}^\pi g(x-y)r(y,t)\,dy\,,
\label{def_dx}
\end{equation}
where $d_0$ is the baseline death rate, $\gamma$ is a constant describing the fraction of resources to be consumed, and $r(y,t)$ denotes the quantity of available resources of type $y$ at time $t$. Let us suppose that the distribution of resources is itself described by a population dynamical equation. To obtain an expression for $r(y,t)$ we assume that the population will grow logistically, but also be depleted as it is consumed by the organisms:
\begin{eqnarray}
\frac{\partial r(y,t)}{\partial t} &=& \alpha r(y,t)
\left( 1 - \frac{r(y,t)}{\kappa} \right) \nonumber \\
&&-\gamma r(y,t) \int^{\pi}_{-\pi} g(y-z) \phi(z,t)\,dz,
\label{eqn_for_r}
\end{eqnarray}
where $\kappa$ is the carrying capacity for the resources and $\alpha$ the growth rate. Let us assume that the resources are sufficiently abundant and quick-growing that we can ignore the possibility of them being driven to extinction due to over-feeding by the organisms (i.e. $\kappa$ and $\alpha$ are both large). In that regime the large size of $r(y,t)$ implies a separation of timescales between the fast dynamics of the resources and comparatively slow changes in the main population. For a given population distribution $\phi$, the resource distribution will quickly approach a value close to 
\begin{equation}
r(y,t)=\kappa\left(1-\frac{\gamma}{\alpha}\int_{-\pi}^\pi g(y-z)\phi(z,t)\,dz\right)\,,
\end{equation}
which, according to our assumptions on the parameters, is a stable positive fixed point of (\ref{eqn_for_r}). Inserting this definition into equation (\ref{def_dx}) yields
\begin{equation}
\begin{split}
&d(x)=d_0-2\pi \gamma\kappa +\frac{2\pi \gamma^2\kappa}{\alpha}\int_{-\pi}^\pi h(x-z)\phi(z,t)\,dz\,,
\end{split}
\end{equation}
where $h$ is given by (\ref{def_h2}). This expression shows how the slowly varying distribution of resources may be integrated out to reveal an effective interaction between individuals, mediated by the competition kernel $h$. In principle the parameters here are free (up to assumptions on size), however, to draw fair comparisons with the case of direct competition it makes sense to choose parameters to match the characteristics of the kernel. This means setting $d_0=2\pi \gamma\kappa$, so that the death rate is zero in the absence of other organisms to compete with, and $\alpha=\gamma d_0$, so that the effective interaction $h$ has the same normalisation as in the case of direct competition.

\section{Calculation of $S$, the invasibility of the population}
\label{appS}

In this appendix we prove the form for $S$ given by Eq.~(\ref{S_result}), from its definition (\ref{def_S}). Any time-dependence will be suppressed throughout.

Using the definition of $\phi$ given in Eq.~(\ref{def_phi}), we have that
\begin{eqnarray*}
\int^{\pi}_{-\pi} s(x) \phi(x)\,dx &=& \frac{1}{K}\,\sum^{N}_{i=1}\,\int^{\pi}_{-\pi} s(x) \delta\left( x - x_i \right)\,dx \nonumber \\
&=& \frac{1}{K}\,\sum^{N}_{i=1}\,s(x_i).
\end{eqnarray*}
This allows us to write $S$ in the alternative way
\begin{eqnarray}
S &=& \int^{\pi}_{-\pi} s(x) \left[ \frac{1}{2\pi} - \frac{K}{N} \phi(x) \right]\,
dx \nonumber \\
&=& \int^{\pi}_{-\pi} \left[ 1 - \int^{\pi}_{-\pi} h(x-y) \phi(y)\,dy \right] 
\left[ \frac{1}{2\pi} - \frac{K}{N} \phi(x) \right]\,dx, \nonumber \\
\label{alt_S}
\end{eqnarray}
using Eq.~(\ref{s_psi}). 

Now,
\[
\int^{\pi}_{-\pi} \left[ \frac{1}{2\pi} - \frac{K}{N} \phi(x) \right]\,dx =
1 - \frac{K}{N}\,\int^{\pi}_{-\pi} \phi(x)\,dx = 0,
\]
using Eq.~(\ref{def_phi}), and
\[
\frac{1}{2\pi}\,\int^{\pi}_{-\pi} dx\,\int^{\pi}_{-\pi} dy\,h(x-y) \phi(y)\,dy 
= \int^{\pi}_{-\pi} dy\,\phi(y) = \frac{N}{K},
\]
again using Eq.~(\ref{def_phi}). Therefore Eq.~(\ref{alt_S}) becomes
\begin{equation}
S = - \frac{N}{K} + \frac{K}{N}\,\int^{\pi}_{-\pi} dx\,\int^{\pi}_{-\pi} dy\,\phi(x)h(x-y)\phi(y).
\label{intermediate_S}
\end{equation}
The double integral in this expression is almost exactly the same as that appearing Eq.~(\ref{Qfour}) for $Q$ --- only $h''$ is replaced by $h$. It may be evaluated in the same way, leading to an analogous expression to Eq.~(\ref{Qone}):
\begin{equation}
S = - \frac{N}{K} + \frac{K}{N} \sum_k \left(\frac{g_k}{2\pi}\right)^{\hspace{-1pt}2}\left| \phi_k \right|^2\,.
\label{S_final}
\end{equation}
We now recall, from integrating over Eq.~(\ref{def_phi}) and from Eq.~(\ref{FT_inverse}), that 
\begin{equation}
\phi_0 = \int^{\pi}_{-\pi} dx\,\phi(x) = \frac{N}{K},
\label{phi_zero}
\end{equation}
and so the $k=0$ term in the sum appearing in Eq.~(\ref{S_final}) is 
\[
\frac{N}{K}\left(\frac{g_0}{2\pi}\right)^{\hspace{-1pt}2} = \frac{N}{K},
\]
since $\int^{\pi}_{-\pi} g(x)\,dx = 2\pi$. Thus this term exactly cancels the
$-N/K$ term in Eq.~(\ref{S_final}) and Eq.~(\ref{S_result}) is proved.

\section{The mesoscopic evolution equation}
\label{appFPE}

In the limit of large carrying capacity, the behaviour of the population is governed by a functional Fokker-Planck equation, first derived in \cite{Rogers2012}:
\begin{equation}
\begin{split}
&\frac{\partial }{\partial t}P(\phi,t)=\\
&-\int^{\pi}_{-\pi} \frac{\delta }{\delta \phi(x)}\left(\mathcal{A}(\phi,x)-\frac{1}{2K}\frac{\delta }{\delta \phi(x)}\mathcal{B}(\phi,x) \right)P(\phi,t)\,dx\,,
\label{FFP}
\end{split}
\end{equation}
where
\begin{equation}
\begin{split}
&\mathcal{A}(\phi,x)=\phi(x)+\mu\nabla^2\phi(x)-\int^{\pi}_{-\pi} \phi(x)\phi(y)h(x-y)\,dy\,,\\
&\mathcal{B}(\phi,x)=\phi(x)+\mu\nabla^2\phi(x)+\int^{\pi}_{-\pi} \phi(x)\phi(y)h(x-y)\,dy\,.
\label{def_AB}
\end{split}
\end{equation}
In this equation, $P$ describes the probability of finding population density $\phi$ at time $t$, while $\mathcal{A}$ and $\mathcal{B}$ capture the deterministic dynamics and demographic noise effects.

Equivalently, the stochastic dynamics of the system may be formulated as a stochastic partial differential equation with statistical properties matching those of the functional Fokker-Planck equation (\ref{FFP}). Following \cite{Gardiner1985}, this formulation is given by
\begin{equation}
\frac{\partial \phi(x,t)}{\partial t} = \mathcal{A}(\phi,x) + 
\frac{\eta(x,t)}{\sqrt{K}}\,,
\label{SDE_gen}
\end{equation}
where $\eta(x,t)$ is a Gaussian white noise with zero mean and correlator 
\begin{equation}
\left\langle \phi(x,t) \phi(x',t') \right\rangle = \mathcal{B}(\phi,x)\,\delta(x-x')\,\delta(t-t'),
\label{correlator}
\end{equation}
understood in the sense of It\={o}. This is precisely equation (\ref{SDE}) from the main text.

As described in the main text (see Eq.~(\ref{def_zeta})), we introduce the new variable $\zeta(x,t) = \phi(x,t) - (1/2\pi)$. To perform the integral in Eq.~(\ref{FFP}) we express the components in terms of Fourier coefficients:
\begin{eqnarray}
\mathcal{A}(\phi,x) &=& -\frac{1}{2\pi}\sum_k\left(\mu k^2+\frac{h_k}{2\pi}\right)\zeta_ke^{ikx} \nonumber \\
&-& \frac{1}{4\pi^2}\sum_{k,l}h_k\zeta_k\zeta_le^{i(k+l)x}\,, \nonumber \\
\mathcal{B}(\phi,x) &=& \,\frac{1}{\pi}+\frac{1}{2\pi}\sum_k\left(2-\mu k^2+\frac{h_k}{2\pi}\right)\zeta_ke^{ikx} \nonumber \\
&+& \frac{1}{4\pi^2}\sum_{k,l}h_k\zeta_k\zeta_le^{i(k+l)x}\,, \nonumber \\
\frac{\delta }{\delta \phi(x)} &=& \sum_ke^{-ikx}\frac{\partial}{\partial \zeta_k}\,.
\end{eqnarray}

\vspace{1cm}

Performing the integral over $x$, we obtain
\begin{equation}
\begin{split}
\frac{\partial P}{\partial t}&=\sum_k\frac{\partial}{\partial \zeta_k} P\left[\left(\mu k^2+\frac{h_k}{2\pi}\right)\zeta_k+\frac{1}{2\pi}\sum_l h_l\zeta_l\zeta_{k-l}\right]\\
&+\frac{1}{2K}\sum_{k,l}\frac{\partial}{\partial \zeta_k}\frac{\partial}{\partial \zeta_l}P\Bigg[\left(2-\mu (k+l)^2+\frac{h_{k+l}}{2\pi}\right)\zeta_{k+l}\\
&\hspace{98pt}+\frac{1}{\pi}+\frac{1}{2\pi}\sum_mh_{m}\zeta_{m}\zeta_{k+l-m} \Bigg]\,,
\end{split}\label{CFP}
\end{equation}
where we use $P$ as a shorthand for $P(\{\zeta_p\},t)$, the distribution of Fourier coefficients of $\zeta(x)$. To calculate the quantities $S$ and $Q$, defined in equations (\ref{def_Q}) and (\ref{def_S}), we are required to make predictions for the behaviour of $|\zeta_k|^2$. We will consider the typical behaviour of the model, as captured by the ensemble average 
\begin{equation}
\big\langle\cdots\big\rangle = \int (\cdots) P(\{\zeta_p\},t)\prod_{l}
d\zeta_l\,.
\label{av}
\end{equation}
Equations of motion for the moments of $P$ are deduced by inserting (\ref{CFP}) into the definition (\ref{av}) and integrating by parts (we assume that $P$ decays exponentially in the tails). Specifically, multiplying (\ref{CFP}) by $\zeta_k$ and integrating both sides yields Eq.~(\ref{m1}) of the main text. Similarly for any pair of integers $k$ and $l$ we find Eq.~(\ref{m2}).

\end{appendix}

%

\end{document}